\documentclass[11pt]{article} 
\usepackage{graphicx}   
\newlength{\dinwidth}
\newlength{\dinmargin}
\begin{document}

\title{The H1 silicon vertex detector}

\begin{figure}[t]
  \centering
  \includegraphics[height=2.5cm,clip=true]{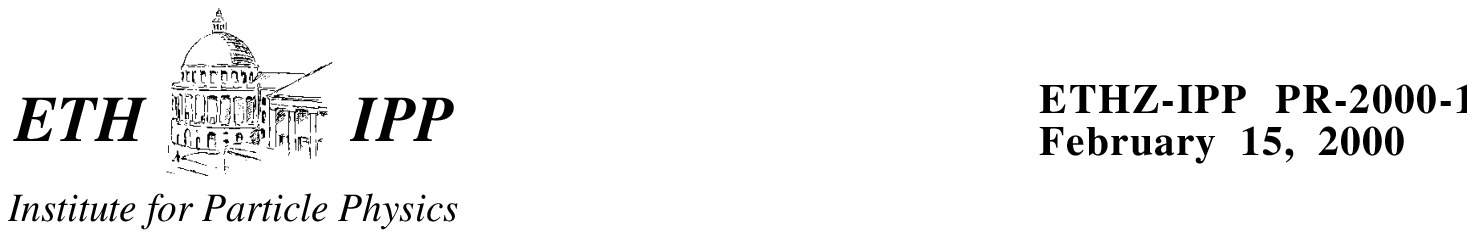}
\end{figure}

\author{D. Pitzl$^{\,a}$, O. Behnke$^{\,d}$, M. Biddulph$^{\,b}$,
K. B\"osiger$^{\,c}$, R. Eichler$^{\,a}$, \\
W. Erdmann$^{\,a}$, K. Gabathuler$^{\,b}$, J. Gassner$^{\,b}$,
W.J. Haynes$^{\,e}$, \\
R. Horisberger$^{\,b}$, M. Kausch$^{\,d}$, M. Lindstr\"om$^{\,f}$,
H. Niggli$^{\,a,b}$, G. Noyes$^{\,e}$, \\
P. Pollet$^{\,a}$, S. Steiner$^{\,c}$, S. Streuli$^{\,a}$, K. Szeker$^{\,a}$,
P. Tru\"ol$^{\,c}$ \\[12pt]
\small $^a$ Institute for Particle Physics, ETH Z\"urich, Switzerland \\
\small $^b$ Paul Scherrer Institute, Villigen, Switzerland \\
\small $^c$ Physics Institute, University Z\"urich, Switzerland \\
\small $^d$ DESY, Hamburg, Germany \\
\small $^e$ Rutherford Appleton Laboratory, Chilton, didcot, UK \\
\small $^f$ Physics Department, Univeristy of Lund, Sweden
}

\maketitle
\begin{abstract}
The design, construction and performance of the H1 silicon vertex detector
is described.
It consists of two cylindrical layers of double sided, double metal
silicon sensors read out by a custom designed analog pipeline chip.
The analog signals are transmitted by optical fibers to a custom
designed ADC board and are reduced on PowerPC processors.
Details of the design and construction are given and
performance figures from the first data taking periods are presented.
\end{abstract}
%
\section{Introduction}
The Central Silicon Tracker (CST) of the H1 experiment at the HERA
electron-proton collider of DESY has been built to provide vertex information
from precision measurements of charged particle tracks
close to the interaction point. It consists of two concentric
cylindrical layers of silicon sensors with two-coordinate readout allowing
the identification of heavy-flavour particles with decay lengths of a few
hundred micrometers \cite{proposal}. The production cross section for
charmed quark pairs at HERA is of order 1\,$\mu$b which offers a rich
field of physics topics \cite{Eichler} that can be exploited once a large
number of charm events are tagged by the vertex detector. In addition, the
production of $b$-quarks can be studied. The $b$ cross section is smaller
by about two orders of magnitude but the longer lifetimes of B-mesons lead
to a more efficient tagging. The bulk of the heavy quarks are produced
close to threshold such that their decay products have an average transverse
momentum around 0.7 GeV/c. The vertex resolution
is dominated by multiple scattering and the amount of material in front of
the second silicon layer must be kept at a minimum.
This led to a design with all readout electronics arranged at the ends
and a central region consisting essentially only of active sensor material.
\par
Space for the installation of the CST was obtained by
reducing the beam pipe radius from 95\,mm to 45\,mm,
which was the minimum radius required to protect the vertex detector from
the direct and backscattered synchrotron radiation emitted by the electron
beam.
\par
The CST has been fully operational since the beginning
of the 1997 running period.
It complements the original central tracking detectors of H1,
which consist of the main jet-cell drift chamber extending from
20.3\,cm to 84.4\,cm in radius,
interspersed by a drift chamber for z-coordinate measurement
between 46\,cm and 48.5\,cm radius,
and an inner z-drift chamber between 17.35\,cm and 20\,cm radius.
A superconducting coil provides a uniform magnetic field of 1.16\,T.
Further details can be found in \cite{H1det}.
Simultaneously to the implementation of the CST the tracking of electrons
scattered at small deflection angles was made possible with the installation
of initially four and, since 1998, eight disks of silicon sensors
in the Backward Silicon Tracker (BST).
The BST \cite{BST} uses the same frontend ASICs and
the same readout electronics as the CST.
\par
In the following section the layout and mechanics of the CST are described.
Section 3 covers the frontend components, i.e. the sensors,
the readout and control chips, the hybrid and the optical link.
The on-line data processing and monitoring of slow control data is covered
in section 4.
The off-line track linking and the alignment procedure
are explained in section 5.
The performance numbers achieved so far are presented in section 6.
%
\section{Layout}
\subsection{Geometry}
\begin{figure}[htb]
  \centering
  \includegraphics[height=12cm,clip=true]{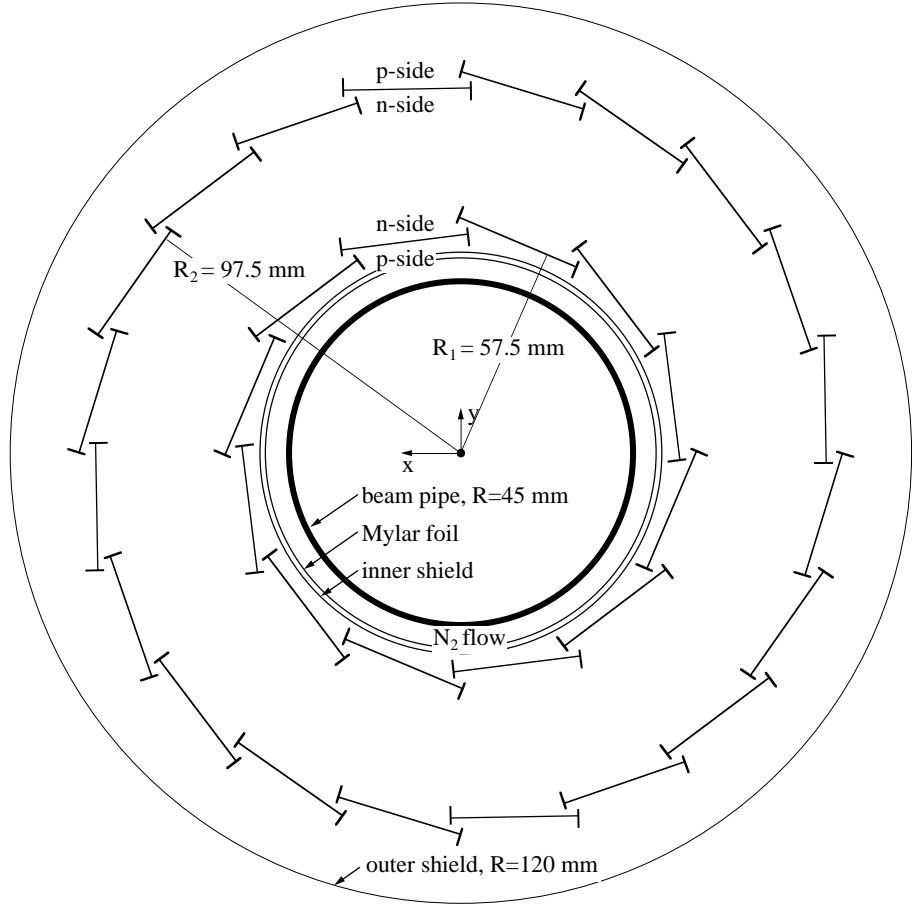}
  \caption[]{CST geometry in the r-$\phi$-plane.}
  \label{r-phi}
\end{figure}
The radial space available for upgrading the H1 experiment
with a vertex detector was limited
on the outside by the first MWPC trigger chamber starting at 15\,cm radius.
On the inside the space was restricted by a beam pipe radius of 4.5\,cm
as required by the synchrotron radiation environment,
and by an additional 7\,mm wide gap
for cooling of the beam pipe with nitrogen gas flowing inside of a Mylar foil.
The beampipe was initially made of aluminium with a wall thickness of 1.7\,mm.
It was replaced in early 1998 by a beam pipe made of 0.15\,mm aluminium
and 0.9\,mm carbon fiber.
\par
\begin{figure}[htb]
  \centering
  \includegraphics[width=\linewidth,clip=true]{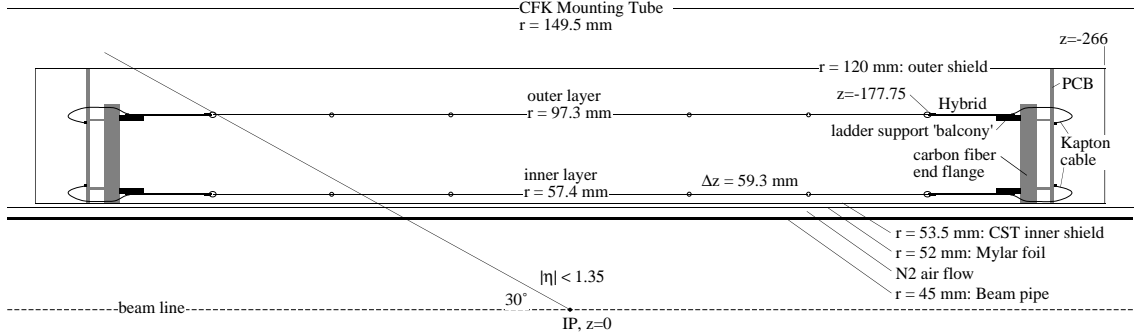}
  \caption[]{Side view of the upper half of the CST.}
  \label{sideview}
\end{figure}

The two layers of the CST are formed from 12 and 20 faces
at radii of 5.75\,cm and 9.75\,cm, respectively, as shown in figure \ref{r-phi}.
One face or 'laddder' consists of six silicon sensors and aluminium nitride
hybrids at each end (see figure \ref{sideview}).
A double layer of carbon fiber strips
with a total thickness of 700\,$\mu$m and a height of 4.4\,mm is glued to the edges.
The carbon fiber strips were specified with a Young's modulus
of at least 400\,000\,N/mm$^2$. The gravitational sag of a full ladder
when supported at the outer ends was measured to be less than 6\,$\mu$m.
\par
The positions of the ladders in a layer are shifted tangentially
to ensure an overlap in r-$\phi$ of adjacent active areas,
which amounts to 1.5\% in the inner layer and 2.1\% in the outer layer.
The active length in $z$ is 35.6\,cm for both layers,
see figure \ref{sideview},
to be compared to the length of the luminous region at HERA 
with an rms width of 10\,cm.
The coverage of the outer layer extends over $\pm1.35$ units
in pseudorapidity for tracks emerging from the origin.
The length is a compromise between rapidity coverage and preamplifier noise
which is proportional to the length.
%
\subsection{Mechanical Frame and Installation}
\begin{figure}[htb]
  \centering
  \includegraphics[height=12cm,clip=true]{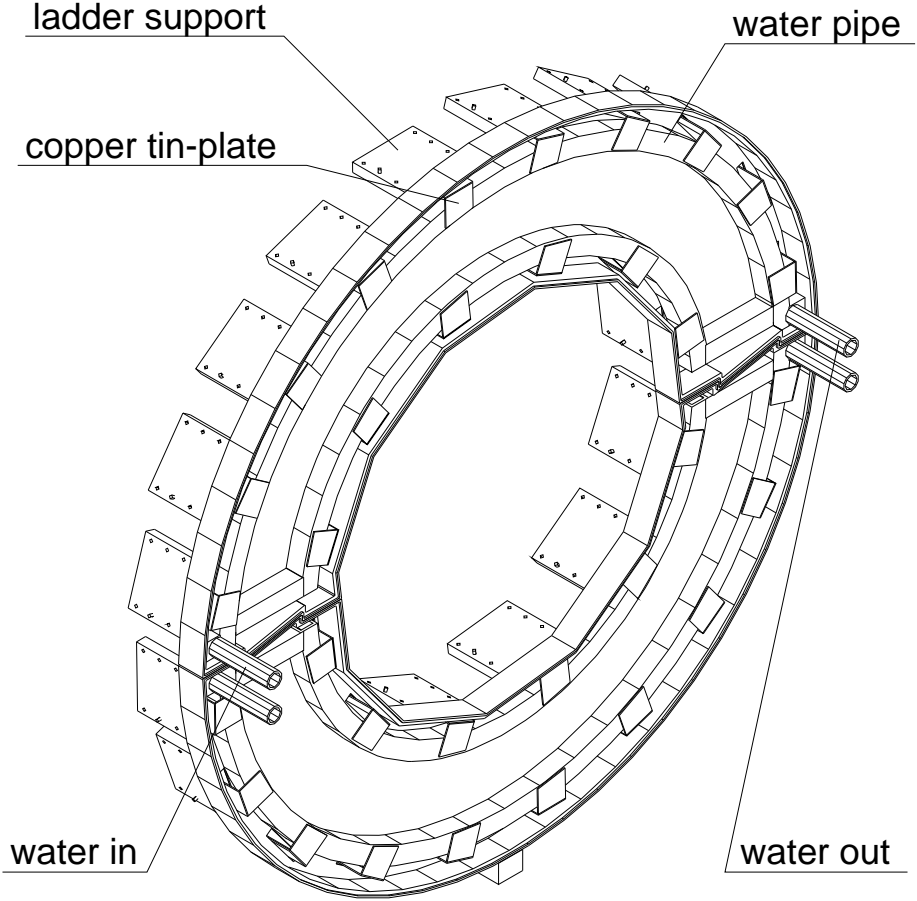}
  \caption[]{Carbon-fiber endflange.
    The cover is removed to display the integrated cooling pipe.
    Copper-tin sheets provide thermal contact to each balcony where the
    hybrids are mounted.}
  \label{end-flange}
\end{figure}

The ladders are mounted on small balconies
extending from carbon fiber endflanges (see figure \ref{end-flange}).
These balconies contain a high precision metal pin used to position the
hybrids at laser-cut holes.
Two small screws on each hybrid are used for fixation.
The carbon fiber endflanges house a circular cooling water pipe for each layer with
copper-tin sheets attached which reach into the balconies and thus provide
thermal contact with the hybrids.
The power dissipation of the CST is 50 W \cite{optical}. 
This power is removed with 12$^o$C cold water at a total flow rate of 2$\ell$/min.
The equilibrium temperature rises from
19$^o$C for the unpowered detector to 28$^o$C during operation.

\begin{figure}[htb]
  \centering
  \includegraphics[width=\linewidth,clip=true]{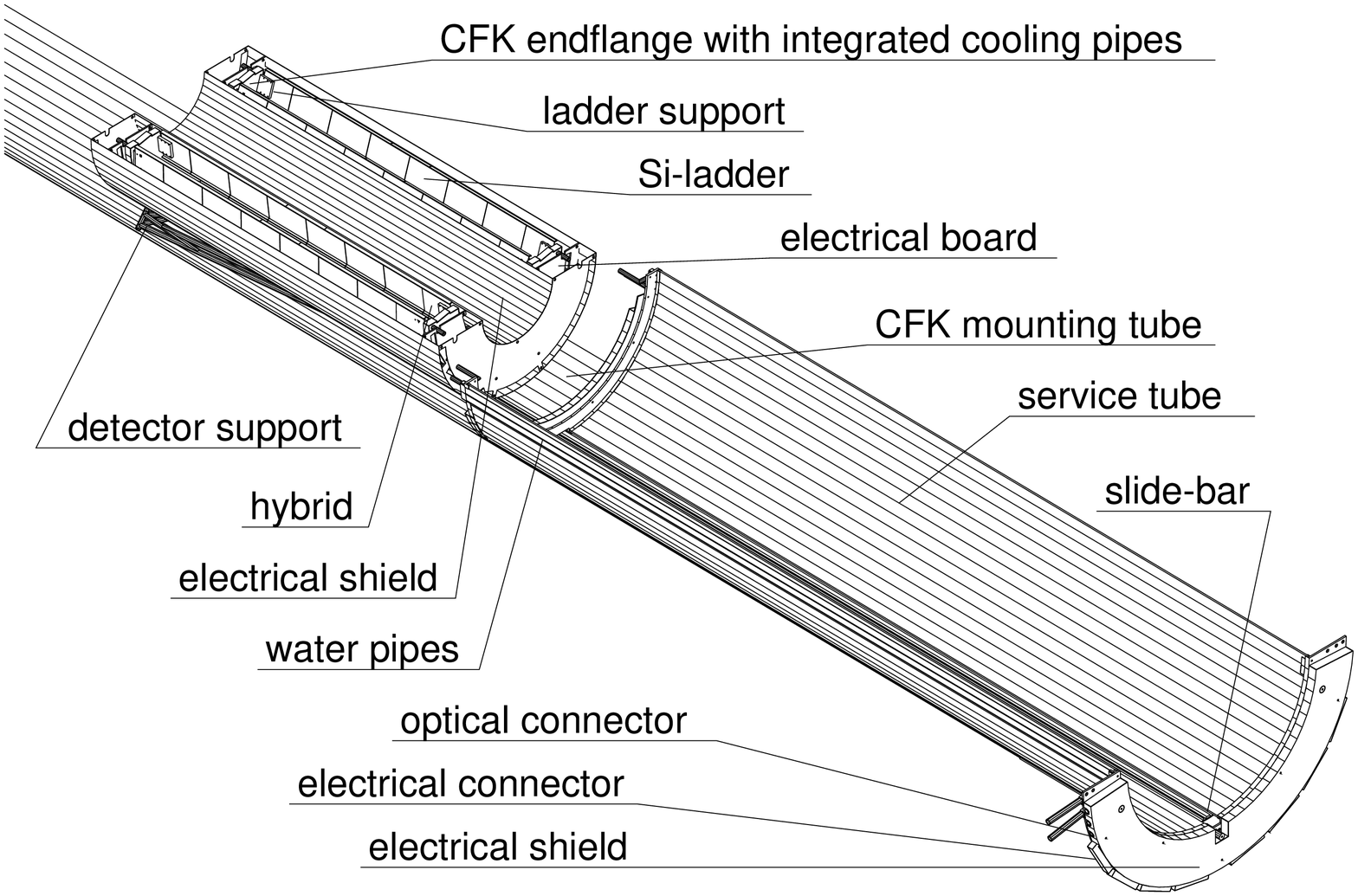}
  \caption[]{View of the lower half of the CST and
    the service tube which surrounds the BST and contains the supply cables,
    the optical fibers and pipes for cooling water.
    Both are supported by the carbon fiber 
support tube which contains
    a carbon fiber sliding rail for insertion.
    The beam pipe is not shown.}
  \label{Adapter}
\end{figure}

The endflanges are split in the horizontal plane (see figure \ref{end-flange}) allowing
for the installation around the beam pipe.
The lower half of the CST rests on three carbon fiber legs in a carbon fiber support tube
(see figure \ref{Adapter}) which
is attached to the innermost
tracking chamber of H1. The upper half of the CST rests on the lower half.

Upon installation the two halves of the CST are first mounted on rail
extensions  around the beam pipe about 2 m from the final position.
A split service tube enclosing power leads, optical fibers and cooling pipes is
equally mounted behind the CST and connections between the CST and the service
tube are made.
Then the CST and the service tube slide on straight carbon fiber rails, 
integrated in the support tube, into the final position, which
is defined by spring-loaded end stops.

The service tube, depicted in figure \ref{Adapter},
has a radial width of only 2\,mm and surrounds the backward silicon tracker
\cite{BST}.
The wall of the service tube is made of a sandwich of 20 $\mu$m
aluminium foil, 2 mm Rohacell \cite{Rohacell} with grooves
for the aluminium power leads and another 20 $\mu$m aluminium foil.
At the edges of the half-shells flat cooling pipes of 2\,mm height are
incorporated.
Cable connectors and voltage regulators are integrated in the service tube
endflange facing away from the CST.

\section{Frontend Components}
\begin{figure}[htb]
  \centering
  \includegraphics[bb=40 90 560 780,height=15cm,clip=true]{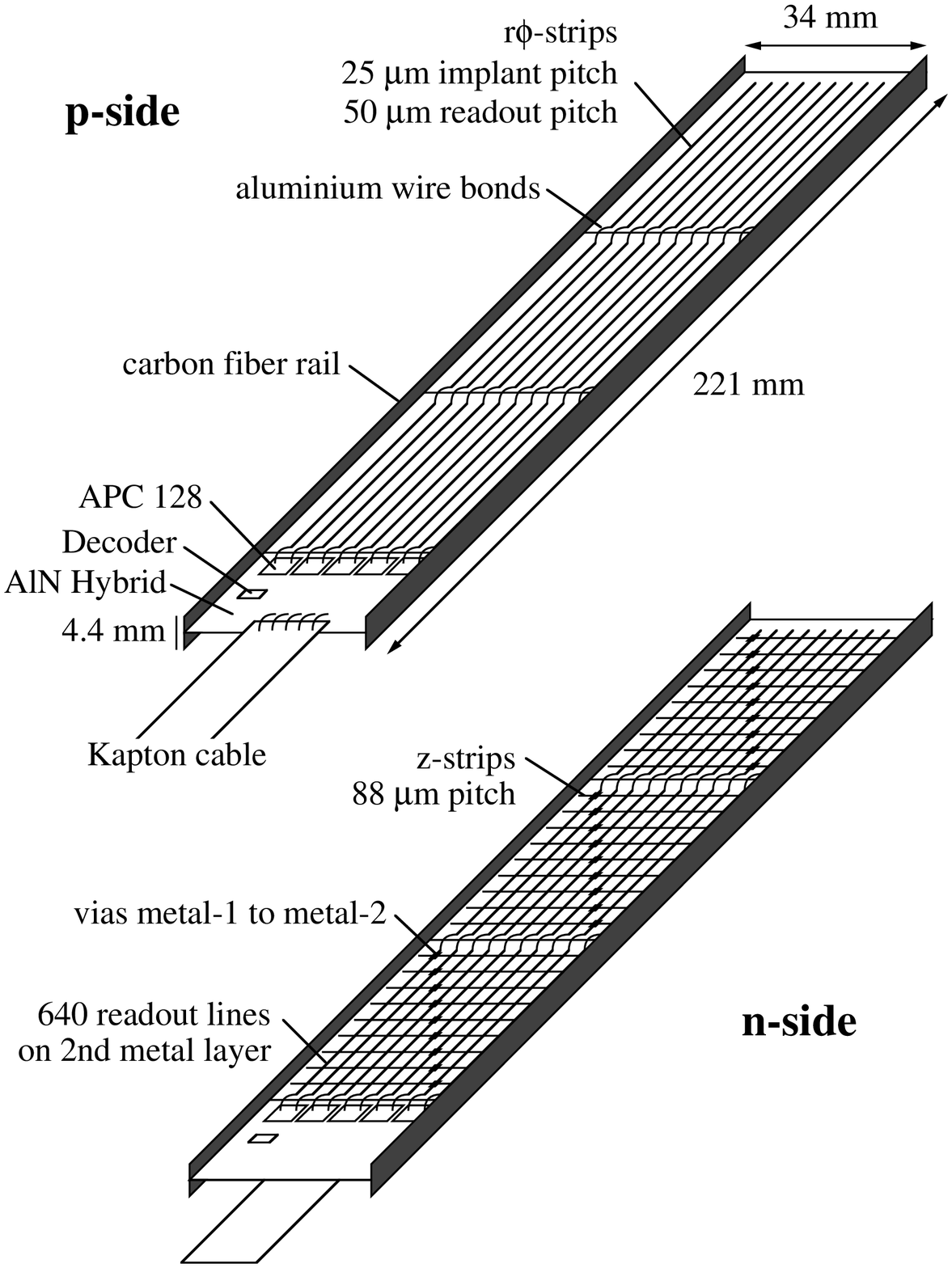}
  \caption[]{Schematic view of a half ladder. The upper 
    part shows the p-side, the lower part the n-side.}
  \label{half-ladder}
\end{figure}

Each ladder consists of two electrical units, called half-ladders.
A half-ladder consists of three silicon sensors of 300\,$\mu$m thickness, 
and a ceramic hybrid of 635\,$\mu$m thickness carrying the
front-end electronics, see figure \ref{half-ladder}.
\par
The silicon sensors have 12\,$\mu$m wide strip implants on both sides.
The strips on the p-side, where holes are collected, are
oriented parallel to the beam direction and have a pitch of 25\,$\mu$m.
Every second strip is read out for a measurement of the
$\phi$-coordinate at a known radius.
The intermediate strips contribute to the signal by capacitive coupling
and improve the position resolution .
The implants on the n-side, where electrons are collected, are rotated by
90$^o$ with respect to the p-side strips and have a pitch of 88\,$\mu$m
for a measurement of the z-coordinate.
Every n-strip is read out by means of a second metal layer
integrated on the sensors.
There are 640 readout lines on each side of a sensor, which are 
daisy-chained by aluminium wire bonds between sensors, and connected to
preamplifier ASICs on the hybrid. This arrangement leads to an effective
strip length of 17.3\,cm on the p-side and to a three-fold ambiguity
for the z-coordinate on the n-side. The insensitive region at
each end of the sensors and a gap of 300\,$\mu$m between the sensors
lead to a coverage in the z-direction of 97\,\%
on the p-side and 95\,\% on the n-side.
In total, the CST contains 64 half-ladders with 192 silicon sensors
and 81\,920 readout channels.
%
\subsection{Silicon Sensors}
High resistivity n-type silicon ($\rho>6$\,k$\Omega$cm)
was obtained as a 100\,mm diameter boule from Wacker Chemitronic
\cite{Wacker}.
Cutting of 300\,$\mu$m wafers and polishing of both sides was
performed by Siltronix \cite{Siltronix}.
The wafer processing was performed at CSEM \cite{CSEM},
where the basic double sided process was extended to provide a second
metal layer over a 5\,$\mu$m thick deposited oxide on the n-side.
The contact vias between metal-1 and metal-2 have a drawn opening of
$12 \times 24$\,$\mu$m$^2$ and proved to be very reliable. Using
contact chain test structures a failure rate of less than $10^{-4}$ was
determined (all CST sensors together contain $1.2\cdot10^5$ vias).
The masks for the 14 layers in this process were designed by the collaboration.

The CST sensors have a full size of $5.9 \times 3.4$\,cm$^2$,
such that two sensors can be produced on a 100\,mm diameter wafer.
The strip implants are DC coupled to the metal-1 layer on both sides.
Early prototypes were AC coupled but showed \cite{Pitzl} a defect rate
for the coupling capacitors on the n-side that led us to resort to DC
coupling.
The intermediate strips on the p-side are biased from a common guard
ring across a punch-through gap covered by a FOXFET gate.
With gate and guard at ground potential and positive bias voltage
applied to the n-side the intermediate strips float at 4\,V,
with a uniformity of about 1\,V on individual detectors and also
between different wafers and different production lots.
We are currently not supplying a dedicated gate voltage,
although this option is available in the cabling scheme.
The active area on the p-side (the junction side) is surrounded by a
multi-ring guard structure with floating gates, that leads to a
gradual increase of the surface potential from 0\,V at the innermost
guard ring to the full bias voltage at the edge. The carbon fiber strips
glued to the sides of the sensors are floating at bias potential.

Each strip on the n-side is surrounded by a narrow ring of p-implant
to provide the necessary interstrip insulation. The n-side can only
be operated at full depletion, which requires between 30\,V and
50\,V for the installed sensors.
Although the strips are DC coupled to the metal-1 layer and all n-side
strips are read out we kept the accumulation channel structure which
provides a high resistance connection to a common guard ring
\cite{Pitzl}. The sensors can then be fully depleted for a measurement
of the total leakage current with only 2 test probe contacts, instead
of having to contact 640 strips on each side.

Detectors with less than 6 $\mu$A of leakage current at 50\,V bias
where selected.
Further tests prior to assembly included sparse measurements of the 
punch-through voltage on the p-side and the conductivity of the 
metal-1 to metal-2 vias on the n-side.
The depletion voltage was determined at several positions on each sensor
by a measurement of the interstrip resistance on the n-side.
Finally, each sensor was scanned under a microscope for shorts or
interrupts in the metallization.
Sensors with more than 6 defective strips on either side were
rejected.
The final yield of accepted sensors was 62\,\% for 9 production lots.

The interstrip capacitance of one strip with respect to its six closest
readout neighbours was measured as 1.5\,pF/cm on the p-side,
for 50\,$\mu$m pitch and 12\,$\mu$m implant width.
On the n-side a value of 19\,pF was measured for the capacitance of
one strip with respect to the other 639 strips on a sensor.
It is dominated by the overlap capacitance between metal-1 and metal-2 lines 
across the 5\,$\mu$m oxide layer.
%
\subsection{Analog Pipeline Chip}
\begin{figure}[htb]
  \centering
  \includegraphics[width=\linewidth,clip=true]{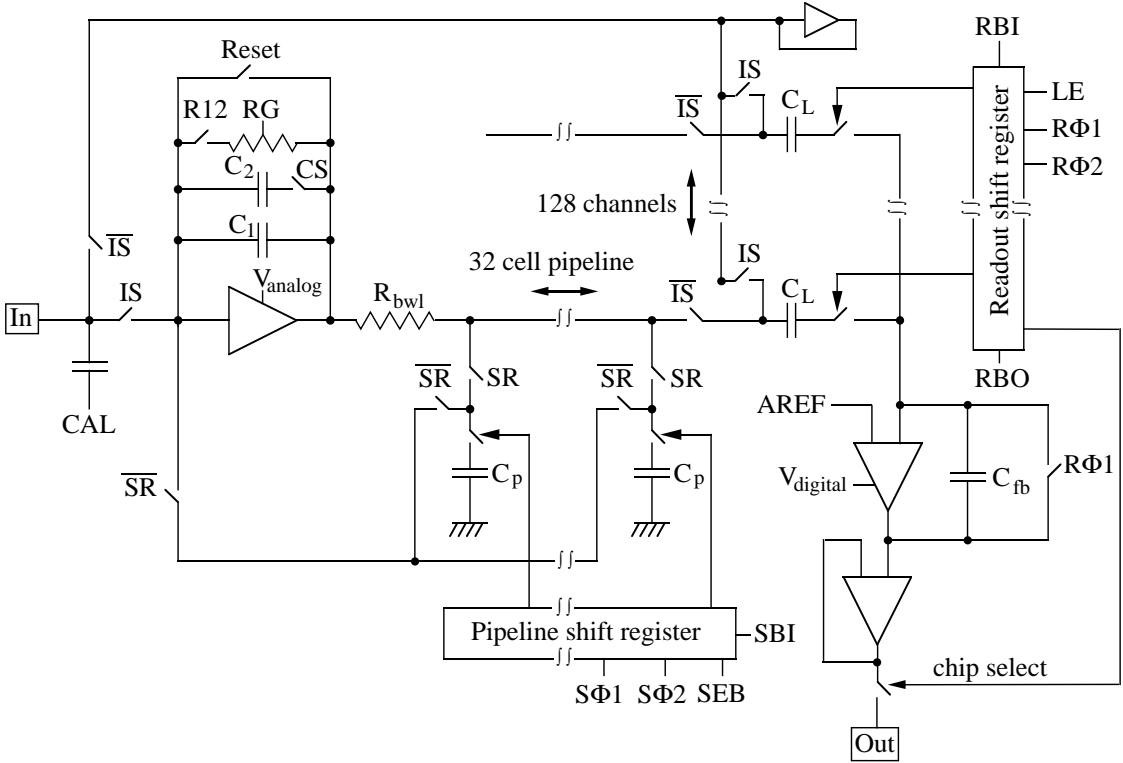}
  \caption[]{Schematic drawing of the APC128 chip showing one channel and
    all peripheral blocks.}
  \label{APCschematic}
\end{figure}
The time between bunch crossings at HERA is 96\,ns
while the H1 level-1 trigger decision arrives after 2.4\,$\mu$s.
Therefore all front-end readout systems have to store the signals from at least
25 beam crossings in a pipeline.
For the H1 silicon detectors an integrated preamplifier and
pipeline chip with multiplexed readout for 128 channels has been
developed \cite{APC} and fabricated in 1.2\,$\mu$m CMOS technology
\cite{Faselec}.
Figure \ref{APCschematic} shows the schematics of the APC128 readout chip.
The various external signals, internal switches and circuit components
are explained in the following sections.
%
\subsubsection{Preamplifier}
The charge sensitive input amplifier consists of a single push-pull
inverter stage which offers minimal noise for a given power dissipation.
The open loop gain is about 150, which, together with a (parasitic)
feedback capacitance C$_1 = 0.45$\,pF, leads to a Miller capacitance
$C_M = (A+1)C_1$ that is not much larger than the input load
capacitance, especially on the n-side.
The equivalent noise charge was measured as
$$  \mbox{ENC} = 700\,e + C_{L} \cdot 50\,e/\mbox{pF} $$
at 0.3\,mW power dissipation and sampling at 10\,MHz \cite{Klaiber}.
The risetime of the amplifier with the detector load was
measured  to be 100\,ns for the p-side and 150\,ns for the n-side.
Due to DC coupling between sensor and chip
the preamplifier must absorb the strip leakage current through
its feedback resistor ($RG$), which is adjustable by an external voltage
and set to a value of about 1\,M$\Omega$.
Consequently the signal decay time is 450\,ns which is sufficient to avoid
pile-up at HERA. Leakage currents of up to several hundred nA per strip
can be tolerated before the preamplifier runs into saturation.

Several switches are used to control the preamplifier.
During data taking the input select switch IS is closed,
connecting the preamplifier input to a strip.
The Reset switch is open and the switch R12 is closed,
activating the feedback resistor RG.
By closing switch CS a second feedback capacitor C$_2$
can be added, which can be used for analog
signal processing \cite{APC} and calibration purposes.
At H1, however, this feature is not used during regular data taking.
The sample/read switch SR connects the preamplifier output to the
switched capacitor analog pipeline.
The preamplifier can be tested by applying a voltage step to the
CAL input. The CAL pulse is reduced internally by about a factor 35
(not shown in figure \ref{APCschematic}). The CAL capacitors of
four neighbouring channels have nominal values of 40, 80, 120 and
160\,fF, which leads to a charge injection corresponding to one to four
minimum ionizing particles in 300\,$\mu$m of silicon for a 3\,V external
test pulse.
%
\subsubsection{Pipeline}
The output voltage of the preamplifier is captured on one of 32
capacitors ($C_p=1$\,pF) that form the analog pipeline for each channel.
The capacitors are cyclically switched under the control of a common
shift register operating at the HERA frequency of 10.4\,MHz.
A sample clock made from two signals ($S\Phi 1$ and $S\Phi 2$),
phase shifted by 50\%, with flat tops and common low periods of at
least several ns is required.
The shift register is cleared by setting both clock signals high and
requires a couple of nanoseconds per cell.
The sample bit ($SBI$) must be refreshed externally every 32 cycles.
%
\subsubsection{Re-read and offset subtraction}
The pipeline is stopped externally at a level-1 trigger signal.
The H1 second level trigger may reject an event after a decision time
of 22\,$\mu$s, upon which the sampling phase is resumed.
An L2 accept decision starts the readout, for which the APC must be put into 
a different mode. First, the input is disconnected from the silicon
sensor by opening the input select switch IS.
This automatically closes the switches $\overline{\mbox{IS}}$
which connects all 128 strips to an extra preamplifier
in auto-feedback configuration to absorb the leakage current
during the readout phase.
Secondly, the sample/read switch SR is opened,
and the switches $\overline{\mbox{SR}}$ are closed,
which disconnects the write lines to the pipeline capacitors
and prepares the read lines.
Thirdly, the reset switch is closed for a few $\mu$s to bring the
preamplifier into a well-defined state.

The APC employs a self-re-reading architecture where the pipeline
capacitors are read back by the same preamplifier that wrote them.
The pipeline cell associated with the triggered event is reached
by advancing the sample bit in the shift register from the stopped position,
refreshing it externally, if necessary.
The sample enable bar switch SEB is open during this phase in order not to
discharge the pipeline capacitors while advancing the sample bit.
The selected capacitor is then read back through the preamplifier
by closing SEB.
The charge stored is amplified by a factor $C_p/C_1 \approx 2.1$
and copied to the latch capacitor C$_L$.
A second and a third sample of the pulse stored in the pipeline
is also read back and added to the charge on C$_L$, which improves
the signal-to-noise ratio by effectively increasing the integration
time.

The latch capacitors are necessary to separate the preamplifier section
of the APC, which operates at a voltage of about 2\,V (V$_{\mathrm{analog}}$),
from the readout section, that operates at 5\,V (V$_{\mathrm{digital}}$).
They also provide intermediate storage of the signals during the
serial readout. Thirdly, they are used to perform an on-chip
pedestal subtraction.
During sampling and up to this point the right plate of the latch
capacitor C$_L$ was connected to the readout amplifier by closing
the latch enable switch LE and permanently filling the readout shift register.
Switch LE is now opened, which captures the signal charge on the right plate.
The left plate is cleared by resetting the preamplifier. The pedestal
is taken from three pipeline capacitors just before the event occurred
and read back with the same procedure as the signal.
With the R12 and Reset switches open, the preamplifier maintains the pedestal
potential, including any shift of the operating point due to
leakage current, at the left plate of C$_L$.
When the readout amplifier is connected to C$_L$ again the difference 
between pedestal and signal is transferred.
%
\subsubsection{Serial readout}
The serial readout is controlled by a shift register which again
requires a two-phased clock signal (R$\Phi 1$ and R$\Phi 2$) and a
readout bit RBI. The right plates of the latch capacitors C$_L$ are
sequentially connected to the readout amplifier having a feedback
capacitance C$_{\mathrm{fb}}$, which provides an amplification of about 10.
A readout speed of 4\,MHz can be reached, if the analog output of the APC
is immediateley followed by a driver amplifier.
For the CST it is limited to 1.6\,MHz by the trace capacitance
on the ceramic hybrid carrying the APC.
The readout of 10 APCs is multiplexed by feeding the readout bit 
appearing at RBO to the RBI input of the next chip.
A chip select mechanism ensures that only one APC at a time connects
to the common readout line.
The full serial readout cycle for 1280 channels requires 1.1\,ms, which
is just sufficient in H1.
%
\subsubsection{Decoder Chip}
The APC requires 13 external signals, of which only the clock and sample
bit signals are fast, while the others change only when switching
from sampling to readout mode.
The number of external clock and control signals that need to be brought
to the front end can be reduced to four by using
a dedicated Decoder chip \cite{optical}.
The desired state of all APC switches is first loaded serially into
registers on the Decoder chip and then applied to the APC.
The fast clock and data signals are passed directly either to the pipeline
or the readout shift register.
Further functionalities have been added to the Decoder Chip:
It can generate a test pulse for the $CAL$ signal at any of
the 32 pipeline buffer positions.
It has a 7-bit DAC which drives a current source for the APC
preamplifiers allowing to define the operating point externally.
Finally, two stabilized and one temperature dependent voltage can
be connected to the readout line, which allows a gain calibration
and temperature monitoring.
The Decoder chip was also fabricated in 1.2\,$\mu$m SACMOS technology
\cite{Faselec}.
%
\subsection{Hybrid and Optical Readout}
Aluminium nitride was chosen as the substrate of the ceramic hybrid
for its excellent heat conductivity $\lambda_{\mbox{AlN}}=160$\,W/Km, compared
to $\lambda_{\mbox{Al$_2$O$_3$}} = 25$\,W/Km for aluminium oxide.
The hybrids have a size of $34 \times 43$\,mm$^2$ and
have two conductor layers on each side.
Connecting vias and holes for fixing screws are cut by laser
\cite{Radeberger}.
One side contains a blank area of $20 \times 16$\,mm$^2$
for heat contact with the mechanical support structure.
Five APCs and one Decoder are mounted on each side of the hybrid
and connected by aluminium wire bonds.
The hybrid carries current sources for the APC preamplifiers,
a voltage reference for gain calibration, a temperature monitor
and drivers for the analog output signal.
The back side of the hybrid, which supplies the n-side of the
silicon sensors, is floating at bias voltage potential.
The digital input signals are transferred across small capacitors
which separate the DC levels.
A thin Kapton cable with 20 lines is glued and wire bonded to the
hybrid and connects to a ring-shaped printed circuit board (endring print)
mounted on the CST endflanges.

The digital control signals and the analog readout are transferred 
by optical fibers \cite{optical} over 34\,m between the detector
and the electronics trailer, which minimizes the amount of
cable material introduced into the center of H1
and prevents electromagnetic interference.
Receivers for a set of four digital control signals are mounted
on four endring prints, each serving one quarter of the CST.
The analog signals are transmitted by a total of 64 LEDs,
which are connected to sockets located on the endring print.
One LED transmits the serial readout of either 1280 p-side channels
or 1280 n-side channels from two neighbouring
half ladders.
The LEDs for the n-side are floating at the bias voltage potential.

\section{Readout and Monitoring}
The frontend system is connected via 34\,m optical fibers and
electrical cables to the readout electronics in the electronics trailer.
The fibers and cables are interrupted twice by connector boards allowing
the installation of the CST and access to other H1 detector components.
%
\subsection{Readout and on-line Data Processing}
\begin{figure}[htb]
  \centering
  \includegraphics[width=\linewidth,clip=true]{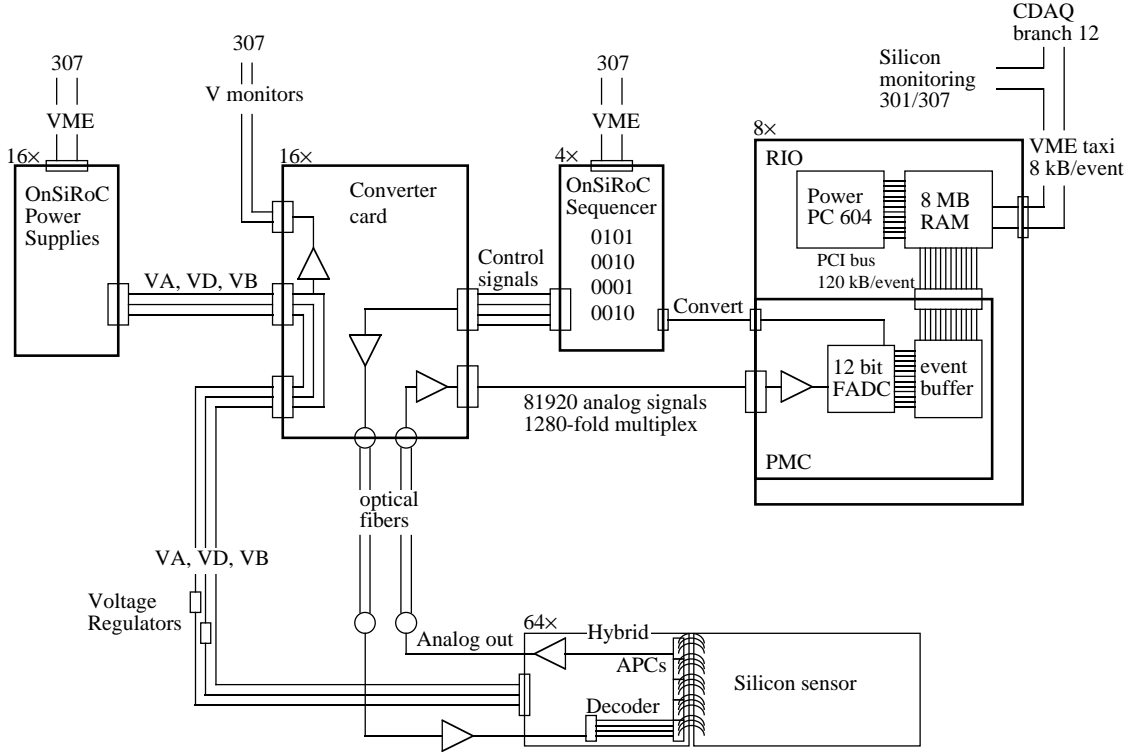}
  \caption[]{Schematic drawing of the CST readout components.}
  \label{readout}
\end{figure}
Figure \ref{readout} shows schematically the components of the readout and
monitoring system.
All electrical and optical leads arrive at a converter card located in the electronics trailer.
It contains LED drivers for the digital control signals and PIN diode
receivers for the analog optical signals.
It also provides passive filtering for the frontend supply voltages and the
detector bias voltage.
The supply voltages are further stabilized by active voltage regulators
placed on the service tube about 1\,m from the detector.
These regulators can be adjusted from the converter cards allowing to
optimize the working points individually for units consisting of
pairs of ladders (four half-ladders).
Finally, the converter cards include circuits for monitoring temperatures,
voltages and detector leakage currents.
If a given temperature limit is exceeded or if the cooling system fails, the converter card
autonomously operates relays switching off the supply voltages to the frontend.

The frontend voltages are generated in VME modules 
called OnSiRoC \cite{Onsiroc}.
The bias voltages are programmable in the range 0\,V to 108\,V.
The OnSiRoC is interfaced to the H1 central trigger and generates
the control sequences required to run the APC128 chips.
A typical sequence occupies 32\,kB in memory and is loaded through VME.
A fast compiler was developed on a Macintosh platform which allows to generate
the sequences from higher level building blocks.

The digitisation of the analog signals is performed on a custom-built
PCI-bus mezzanine card \cite{Bill} using 12 bit FADCs.
The CST creates about 1\,MB of raw data per event, which is
transferred via PCI bus into 8\,MB memories on RIO2 VME cards \cite{CES}.
A hit-finding and zero-suppression algorithm is executed on PowerPC 604
RISC processors operating at 96\,MHz.
The algorithm first determines and subtracts an average baseline for
groups of 128 channels located on individual frontend chips.
The event-to-event variation of this common baseline is comparable to the
single-channel RMS noise.
In a second loop over the data the individual pedestals are subtracted
and hit searching is performed.
A hit is defined as a contiguous group of channels, each with an amplitude
greater than its RMS noise, and with an integrated pulse height of at least
four times the average single channel noise. The hits are copied to an output buffer.
In a third loop the pedestals are updated, using a running average
for each channel and each APC pipeline buffer,
and variances for individual noise determination are accumulated,
except for those channels contributing to a hit.
Further counters are used to identify 'hot' channels which are
included in the noise determination even if they contribute to hits, which
eventually results in a higher calculated RMS noise value with a corresponding
reduction of efficiency.
The hit finding algorithm executes in about 7\,ms with 10240 channels
served by one processor,
while the pedestal updating requires 10\,ms
but is executed only every fourth event.
The formatted hit data are sent via a VME-taxi optical link
to the central data acquisition system of H1.
%
\subsection{Radiation Monitor}
The APC128 chip has been tested for radiation sensitivity in a Co$^{60}$
source.
A single chip can tolerate about 1\,kGy before the analog output saturates
due to internal leakage currents.
This limit is lower and depends on the readout speed when several chips
are daisy-chained.
All other front-end components
have been selected for similar radiation tolerance.

A set of silicon PIN-diodes are attached to the outer shield of the CST
\cite{lund}.
They are continously read out, independent of the H1 data acquisition system.
The counting rate is monitored as a function of time
and displayed in the H1 and HERA control rooms.
Counting rates above a certain threshold require beam tuning or optimisation
of collimator settings.
If the conditions cannot be improved within a few minutes the beams
have to be dumped.
This occurs a few times per year, mainly at the beginning of a running period.
The dose determined by dosimeters attached to the CST was 50\,Gy per year in
1996 and 1997 when HERA stored positrons.
During the electron running in 1998 a dose of up to 250\,Gy was accumulated
which led to severe base-line shifts in the APCs in the inner layer.
In the 1999 shutdown the affected ladders were moved to the outer layer and
the readout ordering was changed to be fully efficient for the 1999-2000
running period.
%
\subsection{Temperature and Leakage Current Monitor}
Temperature dependent solid state current sources (AD590)
are mounted on the CST endflanges.
They are directly monitored in the converter card which
operates relays to cut off all supply voltages to the CST,
should the temperature exceed a value of 60$^o$C.
This hard wired safety circuit is independent of the H1 slow control system.
Furthermore the temperature reading is digitized and displayed by a LabView
application in the control room.

Each hybrid houses a voltage divider driven by a 2.5\,V voltage reference, 
one element being an NTC resistor for temperature measurement.
Furthermore a second reference voltage for gain calibration is derived
from the same reference.
The readout sequence directs the Decoder chip to transfer
these voltages over the analog readout chain at the end of each event readout.
A monitoring program with access to the data stream samples and displays
the temperatures and reference voltages and records their history.
It also provides on-line hit-maps and pulse height distributions for
immediate data quality control.
%
\section{Offline reconstruction}
\subsection{Track Linking}
Tracks from the central tracking chambers are extrapolated to the CST
half-ladders where the search region is limited to five units of the
track extrapolation error.
Ambiguities due to multiple track fit hypotheses in the chambers are resolved
by selecting the best combination of hits in the inner and outer CST layer.
If several tracks cross one half-ladder they are sorted according to their
extrapolation error and the best track is linked first.
Tracks are linked down to a separation of 150\,$\mu$m.

The linking of n-side hits must resolve the three-fold ambiguity created
by the daisy-chained readout with a spacing of 5.93\,cm.
Tracks which have been measured in both  z-chambers have extrapolation
errors below 1\,mm in z and are linked unambiguously.
If only CJC information is available the resolution can be above 1\,cm.
For these cases the linking exploits the correlation between the inner and
outer layer and uses the event vertex as a further constraint.
%
\subsection{CST tracks}
The position and direction of a track can be determined from the
hits in both projections and in both layers of the CST.
Together with the curvature measured in the CJC a so-called CST track
can be defined.
These tracks are used in the CST alignment and they provide a largely
unbiased reference for a re-calibration of the CJC and the z-chambers.
%
\section{Alignment}
In order to profit from the high intrinsic position resolution of the CST
the position of each sensor in space must be known with
comparable precision.
The alignment procedure consists of three steps: An optical survey for the
three sensors on a half-ladder, an internal software alignment of the 64
half-ladders relative to each other and a software alignment of the entire CST 
relative to the rest of the H1 tracking system.
%
\subsection{Optical Survey}
Each half-ladder was surveyed using a microscope and a step-motor controlled
x-y stage with 1\,$\mu$m resolution.
A z-coordinate perpendicular to the sensor plane was measured using
the focal adjustment coupled to a digital micrometer.
Each sensor has 12 alignment marks on the metallization layer whose positions
relative to the strip implants are known from the mask design and
within processing tolerances of less than 3\,$\mu$m.
The survey was analyzed in terms of the relative displacements and rotations
of the three sensors on a half-ladder to an accuracy of 3\,$\mu$m and 0.1\,mrad.
It was observed that the individual sensors are not perfectly flat
but are curved with a sagitta of about 30\,$\mu$m over a diagonal.
A common average curvature is used for all sensors in alignment and
reconstruction.
The original wafers were flat within 5\,$\mu$m after cutting and polishing.
The curvature is probably caused by the thick oxide layer
deposited on the n-side.
%
\subsection{Internal Alignment}
The positions of the ladders are defined by the balconies
on the carbon fiber endflanges.
The mechanical precision of the balconies
and the assembly procedure assure that no forces which may
deform the ladders are exerted.
The placement in space is accurate to a few hundred micrometers.
After applying the alignment corrections from the optical survey
the 64 half-ladders are treated as rigid bodies,
which require 384 alignment parameters.
These are determined in a software alignment procedure
using three sets of tracking data.
%
\subsubsection{Cosmic rays}
\begin{figure}[htb]
  \centering
  \includegraphics[bb=12 360 466 630,height=9cm,clip=true]{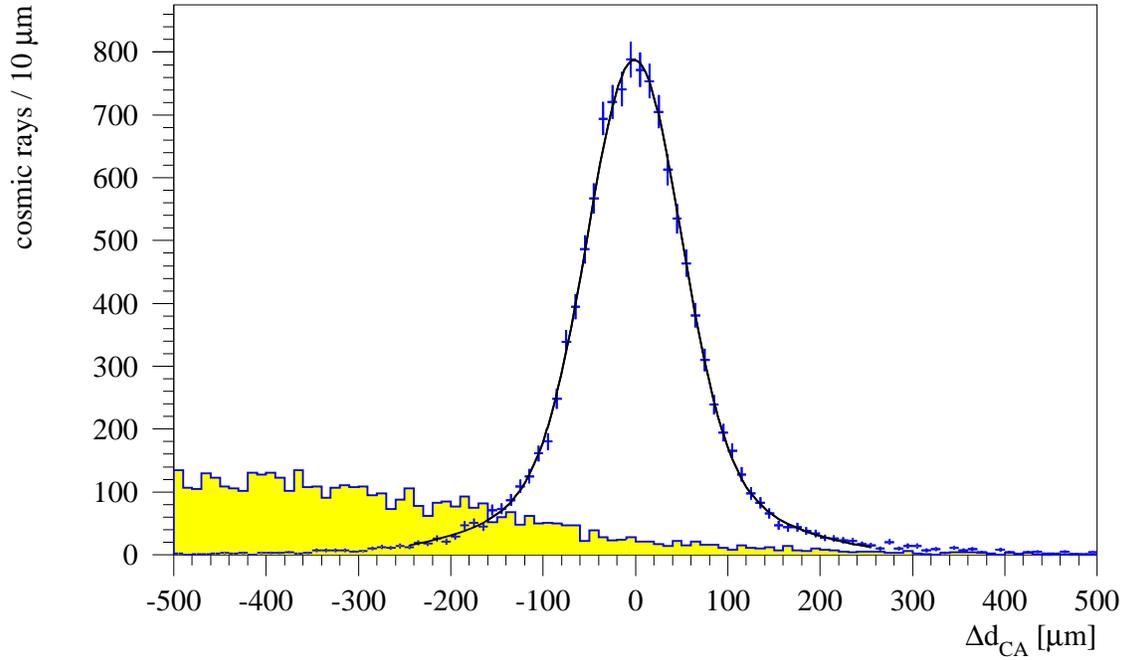}
  \caption[]{Muon miss distance for 4-hit cosmic ray tracks
    with $p_t > 4$\,GeV/c and $|d_{CA}| < 2$\,cm before and after alignment}
  \label{mumiss}
\end{figure}
Cosmic ray data are taken regularly during breaks in the HERA machine
operation.
Penetrating tracks with 4 hits in the CST are selected.
The parameters of the 'upper' and the 'lower' track must agree within errors,
which leads to four constraint equations.
As an example figure \ref{mumiss} shows the distribution of the difference
of the track positions at their closest approach to the origin
of the H1 coordinate system --- the so-called muon miss-distance.
After alignment the standard deviation of the Gaussian is 52\,$\mu$m,
which corresponds to a single-track impact parameter resolution of 38\,$\mu$m
for tracks with a transverse momentum above 4\,GeV/c.
The corresponding impact parameter resolution in the z-projection is 
74\,$\mu m$.
Several million cosmic ray triggers are required for a sufficient
illumination of all half-ladders.
%
\subsubsection{Overlaps}
\begin{figure}[htb]
  \centering
  \includegraphics[bb=16 360 466 630,height=9cm,clip=true]{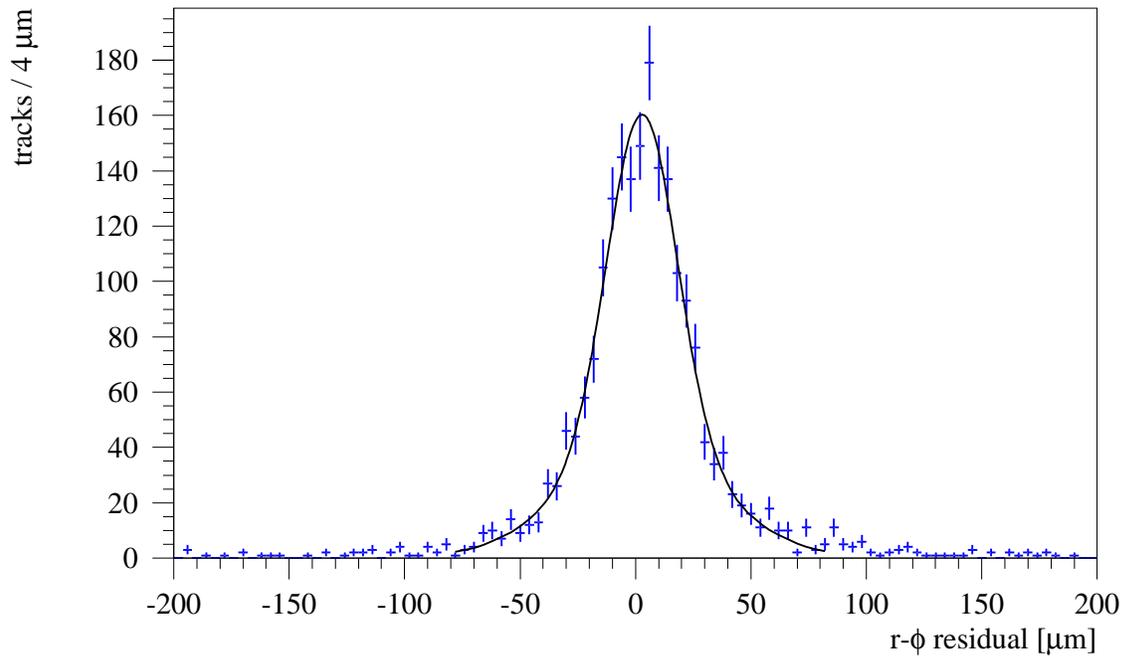}
  \caption[]{Distance between tracks and hits in overlap
    regions in the $r-\phi$ projection after alignment.}
  \label{xres}
\end{figure}
\begin{figure}[htb]
  \centering
  \includegraphics[bb=16 360 466 630,height=9cm,clip=true]{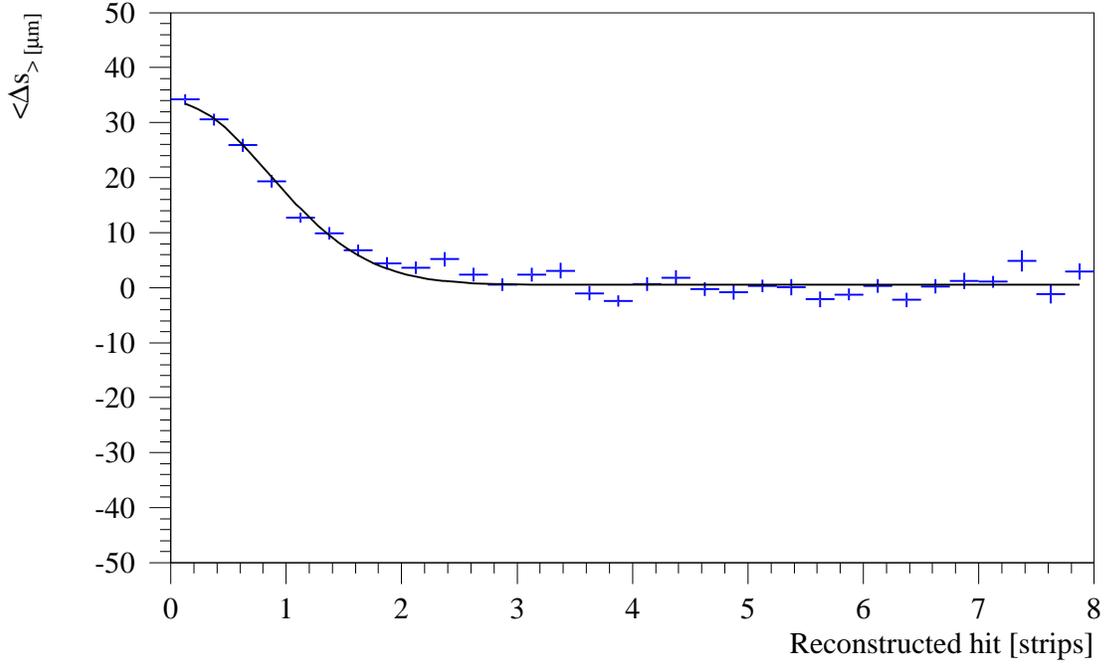}
  \caption[]{Mean overlap residuals versus the reconstructed
    cluster position in units of strip numbers on the p-side
    (50\,$\mu$m pitch). Strip 0 is next to the guard ring.
    The curve is a fit to a semi-Gaussian with a width of 0.85 pitch units.}
  \label{shift}
\end{figure}
\begin{figure}[htb]
  \centering
  \includegraphics[height=9cm,clip=true]{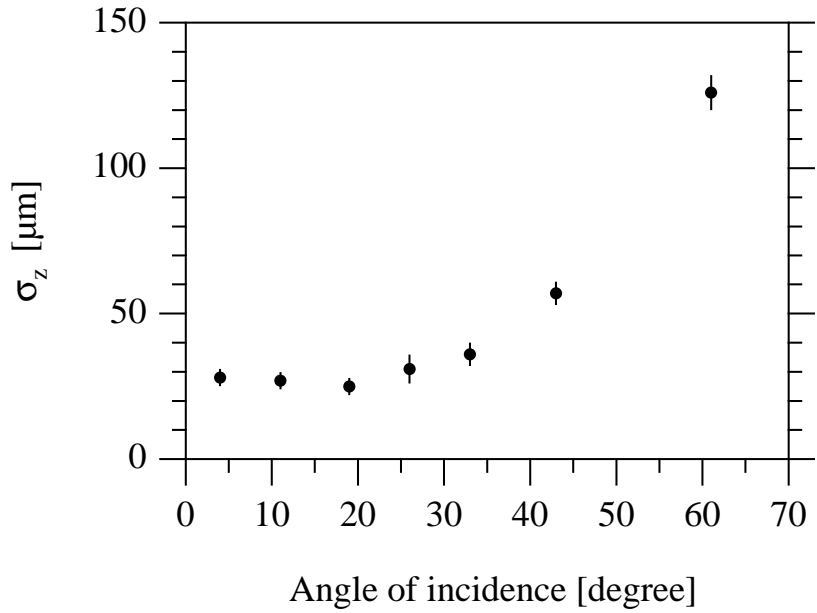}
  \caption[]{Intrinsic resolution in z inferred from overlap
    residuals as a function of incident angle in the r-z projection.}
  \label{zres}
\end{figure}
Cosmic tracks mainly constrain the relative positions of half-ladders in
the inner and outer layer and in the upper and lower half of the CST.
The position of neighbouring half-ladders are constrained by tracks passing
through the overlap regions.
Tracks with 3 hits are selected from normal $ep$ luminosity data and are
used to formulate two constraint equations, one in each readout coordinate.
Two hits are used to define the track and to predict the hit
in the overlap region.
A distribution of residuals in the $r-\phi$ projection is shown
in figure \ref{xres} from which an intrinsic point resolution of 12\,$\mu$m
is inferred.
\par
Close to the guard ring region of the sensors a systematic shift of the
overlap residuals is observed.
In figure \ref{shift} the mean of the residual distribution is shown
as a function of the distance of the reconstructed cluster position
from the guard ring.
The shift is well described by a semi-Gaussian with an amplitude of
33\,$\mu$m and a width of 0.85 pitch units.
The shift is attributed to charge collected on the guard ring.
A correction is made and overlaps on the first two strips are not used
in the alignment procedure.
\par
The angles of incidence do not deviate by more than 22$^o$ from the normal
in the $r-\phi$ projection while much larger angles occur in the $r-z$
projection.
The dependence of the intrinsic z-resolution (measured on the n-side) on
the angle of incidence is shown in figure \ref{zres}. It is well described
by a parabola and reaches a minimum of 22\,$\mu$m at 15$^o$ from normal
incidence \cite{Johannes}.
%
\subsubsection{Vertex Fits}
Multi-track events from $ep$ data are selected and a common 3D event
vertex fit is performed.
The sum of the $\chi^2$ values over several ten thousand events is included
in the overall minimization with respect to the alignment parameters.
This method alone does not lead to a robust estimation of the internal
alignment parameters but together with cosmic rays and overlap tracks
it provides a uniformly distributed track sample of high statistics
that improves the quality of the combined alignment.
%
\subsubsection{Alignment Procedure}
The alignment is performed using the three data sets simultaneously.
A common $\chi^2$ is accumulated and minimized iteratively with respect
to the 384 local alignment parameters.
The sparseness of the corresponding Hessian matrix is exploited for
a fast solution of the linearized equations \cite{blobel}.
Two sets of alignment parameters were determined for 1997,
using alignment data sets taken several months apart.
The parameters are made comparable by applying six overall constraints, that
correspond to a displacements or rotation of the entire CST.
It is found that the internal alignment parameters agree with RMS spreads
of 6\,$\mu$m and 0.1\,mrad.
Compared to the intrinsic silicon resolution this reproducibility and long-term
stability is sufficient.
%
\subsubsection{Global alignment}
The global alignment determines the displacements and tilts
of the entire CST with respect to the CJC and the z-chambers.
Six parameters are determined by minimizing the   
differences between CST and CJC tracks, using $ep$ events and cosmic rays.
%
\section{Performance}
\subsection{Occupancy}
The on-line zero-suppression on average finds 60 p-side clusters and 200 n-side
clusters, corresponding to channel occupancies of 0.8\% and 2\%, respectively.
The higher occupancy on the n-side is due to larger non-Gaussian tails in
the noise distribution. The occupancies are stable in time.
The average number of linked hits is 14 for each projection which represents
the track-related occupancy.
%
\subsection{Signal to Noise Ratio}
\begin{figure}[htb]
  \centering
  \includegraphics[height=7cm,clip=true]{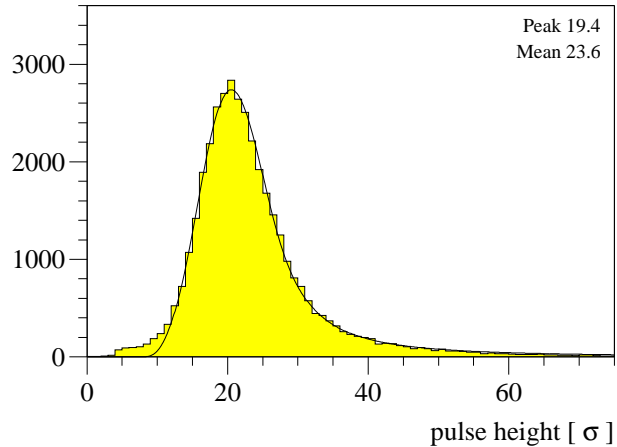}
  \caption[]{p-side cluster pulse height
    divided by the average single channel noise
    for minimum ionizing tracks at vertical incidence.
    A  best-fit Landau curve convoluted with a Gaussian is also shown.}
  \label{landau-p}
\end{figure}
\begin{figure}[htb]
  \centering
  \includegraphics[height=7cm,clip=true]{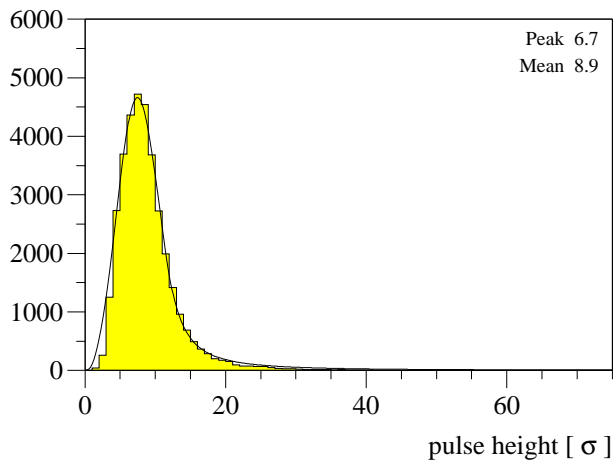}
  \caption[]{n-side cluster pulse height
    divided by the average single channel noise
    for minimum ionizing tracks at vertical incidence.
    The line is a Landau curve convoluted with a Gaussian.}
  \label{landau-n}
\end{figure}
Minimum ionizing particles have a most probable energy loss of 84\,keV in
300\,$\mu$m of silicon, which leads to a signal of about 23\,000
electron-hole pairs.
The thermal noise level is determined by the preamplifier design, its
operating conditions and the detector load capacitance.
For three daisy-chained sensors the capacitance of one strip to
all neighbours amounts to 27\,pF on the p-side and 57\,pF on the n-side,
where the contribution from the double metallization dominates.
The APC is routinely operated in a triple sampling mode and
with a power dissipation of 0.3\,mW per channel.

Figure \ref{landau-p} shows the distribution of cluster pulse heights
divided by the average single-channel noise for cosmic muon tracks,
normalized to vertical incidence.
The shape is well described by a Landau energy loss distribution
with a most probable signal-to-noise ratio of 19 for the p-side and 6.7 for
the n-side, see figure \ref{landau-n}.
The difference is due to the strip capacitance loading the preamplifier
which is a factor of two larger on the n-side, and due to the incomplete
charge amplification caused by the limited gain of the preamplifier.
%
\subsection{Efficiency}
\begin{figure}[htb]
  \centering
  \includegraphics[height=9cm,clip=true]{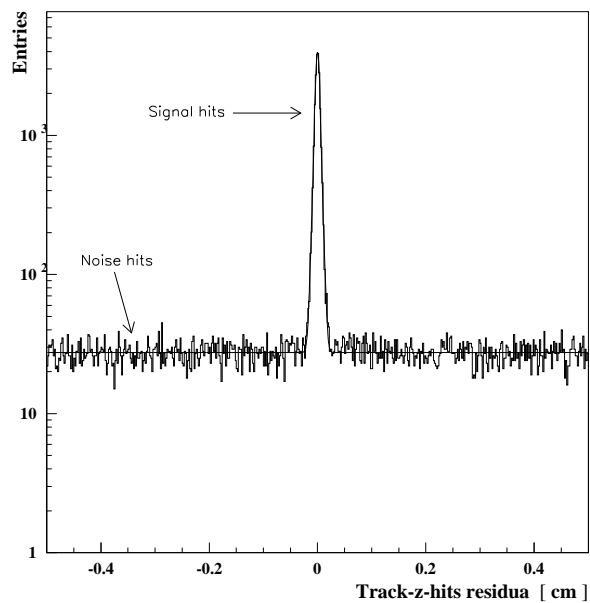}
  \caption[]{Distance between cosmic track intersect
    points and all hits on a half-ladder in the z-projection.}
  \label{tzhr}
\end{figure}

\begin{figure}[htb]
  \centering
  \includegraphics[height=9cm,clip=true]{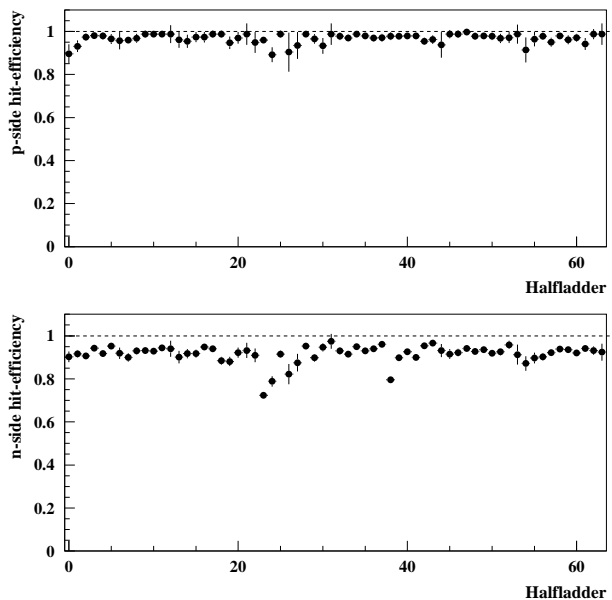}
  \caption[]{Hit efficiencies for p-side (top) and
    n-side (bottom) for all 64 half-ladders as measured from cosmic ray tracks}
  \label{hit-eff}
\end{figure}

The CST hit efficiencies are most accurately determined with
cosmic tracks passing through four CST half-ladders.
Using three linked hits and the curvature from the CJC
the track parameters are determined in a fit and the intersection with
the fourth half-ladder is calculated.
Figure \ref{tzhr}
shows for a sample of 20000 muon tracks with transverse momentum
above 2\,GeV the distance between the intersection point and all 
hits in the test layer in the z-projection.
The central peak at zero contains the signal hits while the noise
hits create a flat background distribution.
The central peak can be described by two gaussians with widths
of 33\,$\mu$m and 64\,$\mu$m for test half-ladders in the inner and outer
layer, respectively.
By comparing the number of hits in the peak
with the number of passing tracks one can determine the 
hit-efficiencies.
Fig. \ref{hit-eff} shows the results for p- and n-side hits
for all 64 half-ladders.
Besides some fluctuations, which can be associated
with specific hardware problems for the selected data runs, 
the efficiencies are in agreement with being the same for all half-ladders.
For the p-side the average efficiency is 97\%, while it
is 92\% for the n-side.
The inefficiencies is caused by silicon defects, dead or noisy
readout channels, the hit finding algorithm and the linking procedure.
The lower efficiency for n-side is due to the lower signal-to-noise
ratio.
%
\subsection{Beam Line Reconstruction}
\begin{figure}[htb]
  \centering
  \includegraphics[bb=55 444 511 755,height=9cm,clip=true]{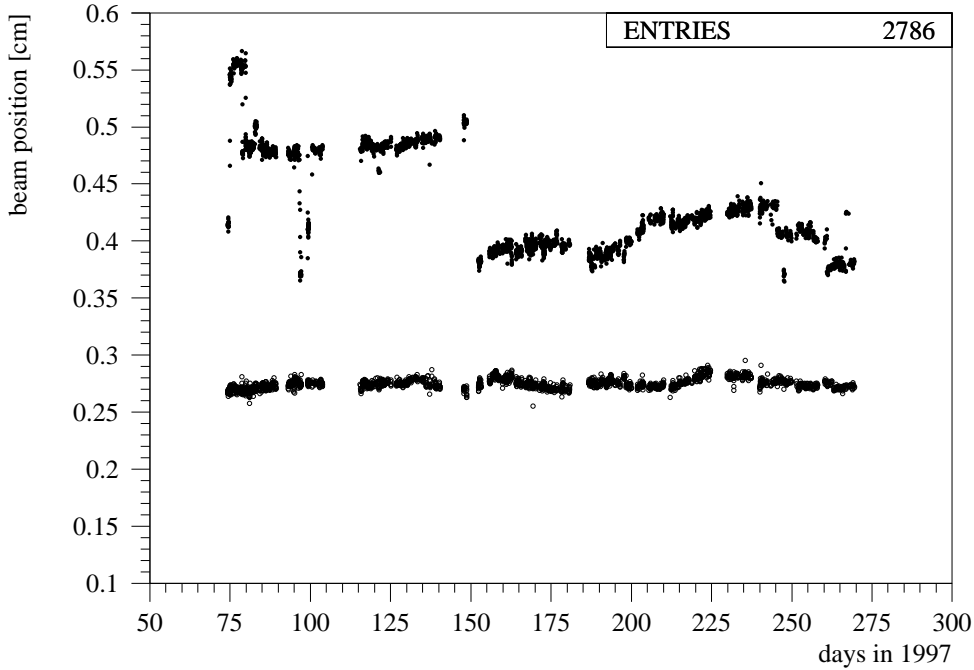}
  \caption[]{HERA beam position during 1997 as determined
    by the CST. The lower band of symbols shows the vertical beam position
    (stable at 0.27\,cm),
    the upper band with a step around day 150 shows the horizontal beam
    position.}
  \label{runvertex}
\end{figure}
A precise knowledge of the beam position as a function of time is required
for many decay-length or impact parameter studies.
The beam position and tilt is determined by accumulating CST tracks over
typically 30 minutes and minimizing the closest approach to a line in space.
Figure \ref{runvertex} shows the horizontal and vertical beam position 
determined for the 1997 luminosity period.
The horizontal beam movements reflect adjustments to the HERA optics.

\begin{figure}[htb]
  \centering
  \includegraphics[bb=16 360 466 630,height=9cm,clip=true]{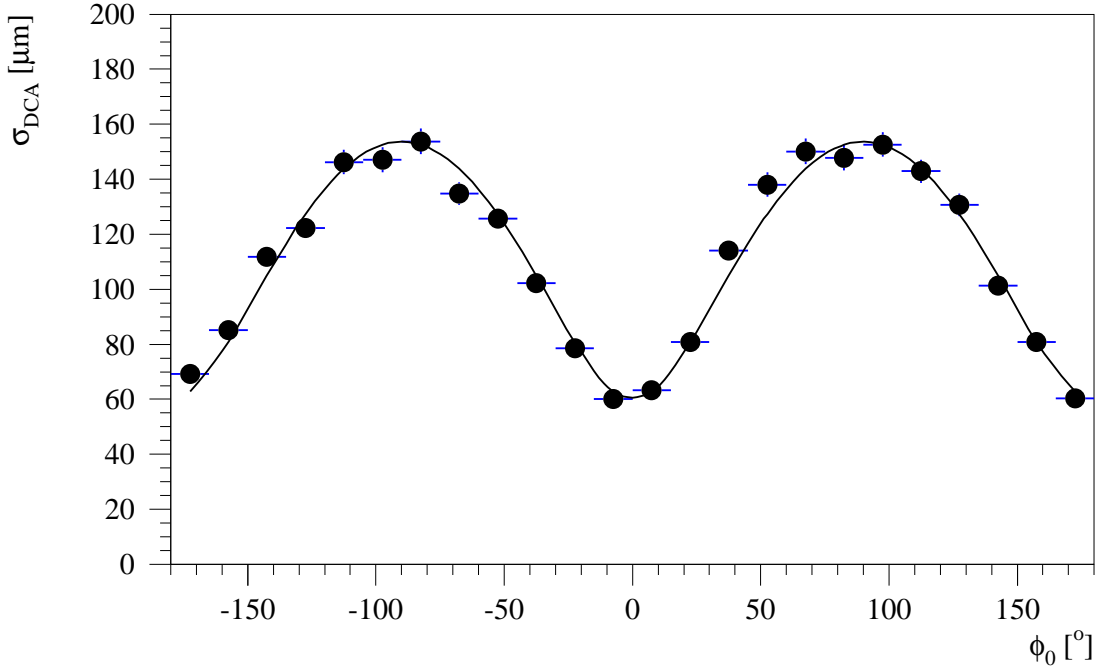}
  \caption[]{Width of the CST impact parameter distribution
    versus the track direction $\phi_0$ around the beam spot
    for transverse momenta above 4\,GeV/c. The curve is discussed in the text.}
  \label{beamspot}
\end{figure}

The remaining distribution of the closest approach to the beam line ($d_{CA}$)
has a central Gaussian part with
contributions from the CST intrinsic resolution, from multiple scattering
in the beam pipe and the first silicon layer and from the beam spot size.
The decays of long-lived particles contribute to the non-Gaussian tails.
From the HERA machine optics an elliptical beam spot with a 
horizontal-to-vertical aspect ratio of 5 to 1 is expected.
This allows to separate the different contributions by measuring the
width of the central Gaussian of the $d_{CA}$ distribution as a function
of the track direction around the beam.
The result is shown in figure \ref{beamspot} for tracks with high
momentum where the multiple scattering contribution can be neglected.
A fit of the form
$$ \sigma^2 = \sigma_0^2 + \sigma_x^2 \sin^2\phi + \sigma_y^2 \cos^2\phi $$
is used to extract the CST intrinsic $d_{CA}$ resolution of $\sigma_0=54\,\mu$m
and a horizontal beam spot size of $\sigma_x=155\,\mu$m, which agrees
with the HERA optics.
A ratio $\sigma_y/\sigma_x=1/5$ as given by the optics
was assumed in the fit to unfold the CST intrinsic resolution.
%
\subsection{Impact Parameter Resolution}
\begin{figure}[htb]
  \centering
  \includegraphics[bb=16 360 466 630,height=9cm,clip=true]{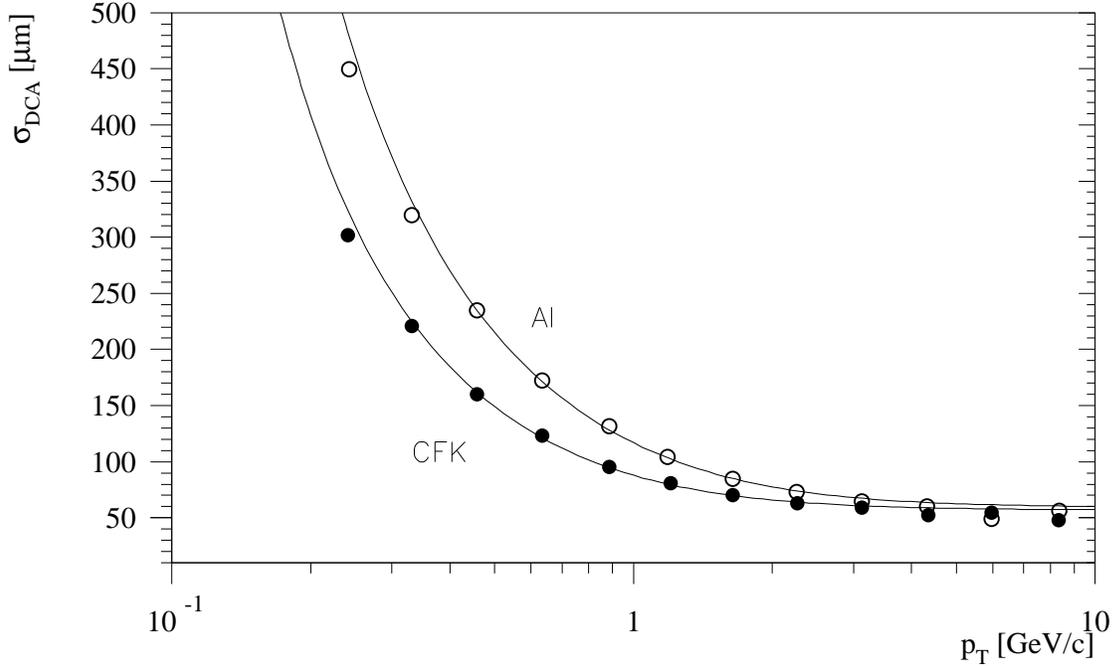}
  \caption[]{CST impact parameter resolution as a function of
    transverse momentum for horizontal tracks (within $\pm15^o$).
    The open symbols are from 1997 (Al beam pipe), the filled symbols from 1999
    (carbon fiber beam pipe). The curves are discussed in the text.}
  \label{dcares}
\end{figure}
The multiple scattering contribution to the width of the $d_{CA}$ can
be measured as a function of momentum by unfolding the contribution of
the beam spot size. This contribution is minimal for horizontal tracks.
The result is shown in figure \ref{dcares} for data from 1997 and from 1999.
A fit according to
$$ \sigma^2 = \sigma_0^2 + (A/p_t)^2$$
leads to asymptotic values $\sigma_0$ of $57\,\mu$\,m and $59\,\mu$m for
the two years while the
parameter $A \sim \sqrt{d/X_0}$ improves by a factor 1.55,
as expected for the change from an aluminium beam pipe ($d=1.9\%\,X_0$)
to a carbon fiber beam pipe ($d=0.6\%\,X_0$), when adding the constant
contribution of $d=0.6\%\,X_0$ the first silicon layer and the CSt inner
shield.
\section{Summary}
The H1 silicon vertex detector CST has been operated successfully at HERA
since the beginning of 1997.
The sensors, the readout electronics and the optical signal transmission are
functioning reliably and efficienctly.
A point resolution of 12\,$\mu$m with a 
signal-to-noise ratio of 19 has been achieved for the $r-\phi$ coordinate,
while the minimal point resolution in $z$ is 22\,$\mu$m with a signal-to-noise
ratio of 7. An impact parameter resolution of 37\,$\mu$m in the $r-\phi$ plane
has been achieved for high momentum tracks, which opens a wide range of
physics topics in the field of heavy quark production in electron-proton
collisions.
\end{document}